\documentclass[twocolumn]{aastex631}

\usepackage{booktabs}
\usepackage{amsmath}
\usepackage{multirow}
\usepackage{array}
\usepackage{booktabs}
\usepackage{CJK}
\usepackage{threeparttable}
\usepackage{pifont}
\usepackage{enumitem} 

\usepackage{stackengine}
\newcommand\xrowht[2][0]{\addstackgap[.5\dimexpr#2\relax]{\vphantom{#1}}}

\shorttitle{AASTeX v6.31 Sample article}
\shortauthors{Wang et al.}

\graphicspath{{./}{figures/}}

\begin{document}
\begin{CJK*}{UTF8}{gbsn}
\title{Density profile of ambient circumnuclear medium in Seyfert 1 galaxies}
\correspondingauthor{Zhicheng He}
\email{zcho@ustc.edu.cn}

\author[0000-0002-1010-7763]{Yijun Wang (王倚君)}
\affiliation{Department of Astronomy, Nanjing University, Nanjing 210093, China}
\affiliation{Key Laboratory of Modern Astronomy and Astrophysics 
(Nanjing University), Ministry of Education, Nanjing 210093, China}
\affiliation{CAS Key Laboratory for Research in Galaxies and Cosmology, 
Department of Astronomy, 
University of Science and Technology of China, Hefei 230026, China}
\affiliation{School of Astronomy and Space Science, 
University of Science and Technology of China, Hefei 230026, China}

\author[0000-0003-3667-1060]{Zhicheng He (何志成)}
\affiliation{CAS Key Laboratory for Research in Galaxies and Cosmology, 
Department of Astronomy, 
University of Science and Technology of China, Hefei 230026, China}
\affiliation{School of Astronomy and Space Science, 
University of Science and Technology of China, Hefei 230026, China}

\author{Junjie Mao (毛俊捷)}
\affiliation{Department of Physical, Hiroshima University, 
1-3-1 Kagamiyama, HigashiHiroshima, Hiroshima 739-8526, Japan}
\affiliation{SRON Netherlands Institute for Space Research, 
Niels Bohrweg 4, 2333 CA Leiden, The Netherlands}

\author{Jelle Kaastra}
\affiliation{SRON Netherlands Institute for Space Research, 
Niels Bohrweg 4, 2333 CA Leiden, The Netherlands}
\affiliation{Leiden Observatory, Leiden University, 
Niels Bohrweg 2, 2300 RA Leiden, The Netherlands}

\author[0000-0002-1935-8104]{Yongquan Xue (薛永泉)}
\affiliation{CAS Key Laboratory for Research in Galaxies and Cosmology, 
Department of Astronomy, 
University of Science and Technology of China, Hefei 230026, China}
\affiliation{School of Astronomy and Space Science, 
University of Science and Technology of China, Hefei 230026, China}

\author{Missagh Mehdipour}
\affiliation{Space Telescope Science Institute, 3700 San Martin Drive, 
Baltimore, MD 21218, USA}

\begin{abstract}

The shape of the ambient circumnuclear medium (ACM) density profile 
can probe the history of accretion onto the central supermassive black hole 
in galaxies and the circumnuclear environment. However, due to the 
limitation of the instrument resolution, the density profiles of the
ACM for most of galaxies remain largely unknown. In this work, we propose a novel 
method to measure the ACM density profile of active galactic nucleus (AGN) by the equilibrium between 
the radiation pressure on the warm absorbers 
(WAs, a type of AGN outflows) and the drag pressure from the ACM. 
We study the correlation between the outflow velocity and 
ionization parameter of WAs in each of the five Seyfert 1 galaxies 
(NGC~3227, NGC~3783, NGC~4051, NGC~4593, and NGC~5548), 
inferring that the density profile of the ACM is between 
$n \propto r^{-1.7}$ and $n \propto r^{-2.15}$ ($n$ is number density and $r$ is distance) 
from 0.01 pc to pc scales in these five AGNs. 
Our results indicate that the ACM density profile in Seyfert 1 galaxies is steeper than the prediction by the 
spherically symmetric Bondi accretion model and the simulated results of 
the hot accretion flow, but more in line with the prediction by the standard thin disk model.

\end{abstract}

\keywords{Galaxies: Seyfert (1447) --- Galaxies: nuclei (609) --- Galaxies: ISM (847) --- X-rays: galaxies (1822)
--- Accretion, accretion disks (562)}

\section{Introduction} \label{sec:intro}

The ambient circumnuclear medium (ACM)
in the center of galaxies can probe the accretion history of the 
central supermassive black hole (SMBH) in galaxies.
Different accretion models correspond to different density profiles of the ACM.
The classical Bondi accretion \citep[spherically symmetrical accretion;][]{Bondi1952} 
predicts that the density profile of the accretion flow is
$n\propto r^{-1.5}$ ($n$ is number density and $r$ is distance) within the Bondi radius,
and is a constant at larger radii \citep{Frank2002}.
The density profile of the hot accretion flow, 
such as advection-dominated accretion flows 
\citep[ADAFs;][for a review]{Narayan1994,Yuan2014},
is between $n\propto r^{-0.5}$ and $n\propto r^{-1}$ according to simulations
\citep{Yuan2012}.
The theory of the standard cold, thin accretion disk \citep{Shakura1973} 
predicts that the density profile of the accretion 
flow is $n\propto r^{-15/8}$ \citep{Frank2002}.
The multi-wavelength observations toward the center of the Milky Way (MW)
indicated that the density profile of the ACM is $n\propto r^{-1}$ at 
several hundred Schwarzschild radii \citep{Gillessen2019},
and $n\propto r^{-1.5}$ in the hot gas halo at the kpc scale \citep{Miller2015}.
{\it Chandra} X-ray observations toward the center of M87 and NGC~3115 show that 
the density profiles of their ACM are $n\propto r^{-1}$ within 
the Bondi radius \citep{Russell2015,Wong2011}.
However, due to the limitation of the instrument resolution,
the density profiles of the ACM for most galaxies are still unknown.
One way to infer the density profile of the ACM
is through fitting the spectral energy distribution of tidal disruption events 
(TDEs; a star disrupted by the tidal forces from the SMBH),
which can trace the interaction process between the outflows 
from TDEs and the ACM \citep{Alexander2016,Eftekhari2018,Anderson2020,Alexander2020}. 
However, TDEs are only detected in a small number of galaxies and 
are difficult to be identified in active galactic nuclei (AGNs) \citep{Gezari2021}.
Besides, for AGNs, the emission from the accretion disk or jet will overshadow
the emission from the interaction between outflows and ACM at small scales.
In this work, we propose a novel way to estimate the density profile of the 
ACM in AGNs.

AGNs usually play an important role in forming and 
driving outflows which might further affect the 
star formation of their host galaxies \citep[][for a review]{He2019,Chen2022,King2015}. 
These outflows might interact with the ACM.
Warm absorbers (WAs) are part of AGN ionized outflows \citep[e.g.,][]{Laha2014}, 
which are detected in roughly half of nearby AGNs \citep[e.g.,][]{Reynolds1997,Kaastra2000,Tombesi2013}.
WAs usually consist of several ionization phases \citep[e.g.,][]{Laha2014} 
and are located from the accretion disk to the narrow-line region
\citep[e.g.,][]{Reynolds1995,Elvis2000,Blustin2005}.
WAs have the outflowing velocities up to a few thousand of km s$^{-1}$ 
\citep[e.g.,][]{Kaastra2000,Ebrero2013}, 
and are considered to be driven by radiation pressure \citep[e.g.,][]{Proga2004}, 
magnetic forces \citep[e.g.,][]{Blandford1982,Fukumura2010}, 
or thermal pressure \citep[e.g.,][]{Begelman1983,Mizumoto2019}.

\begin{figure*}[ht!]
\centering
\includegraphics[width=0.9\linewidth, clip]{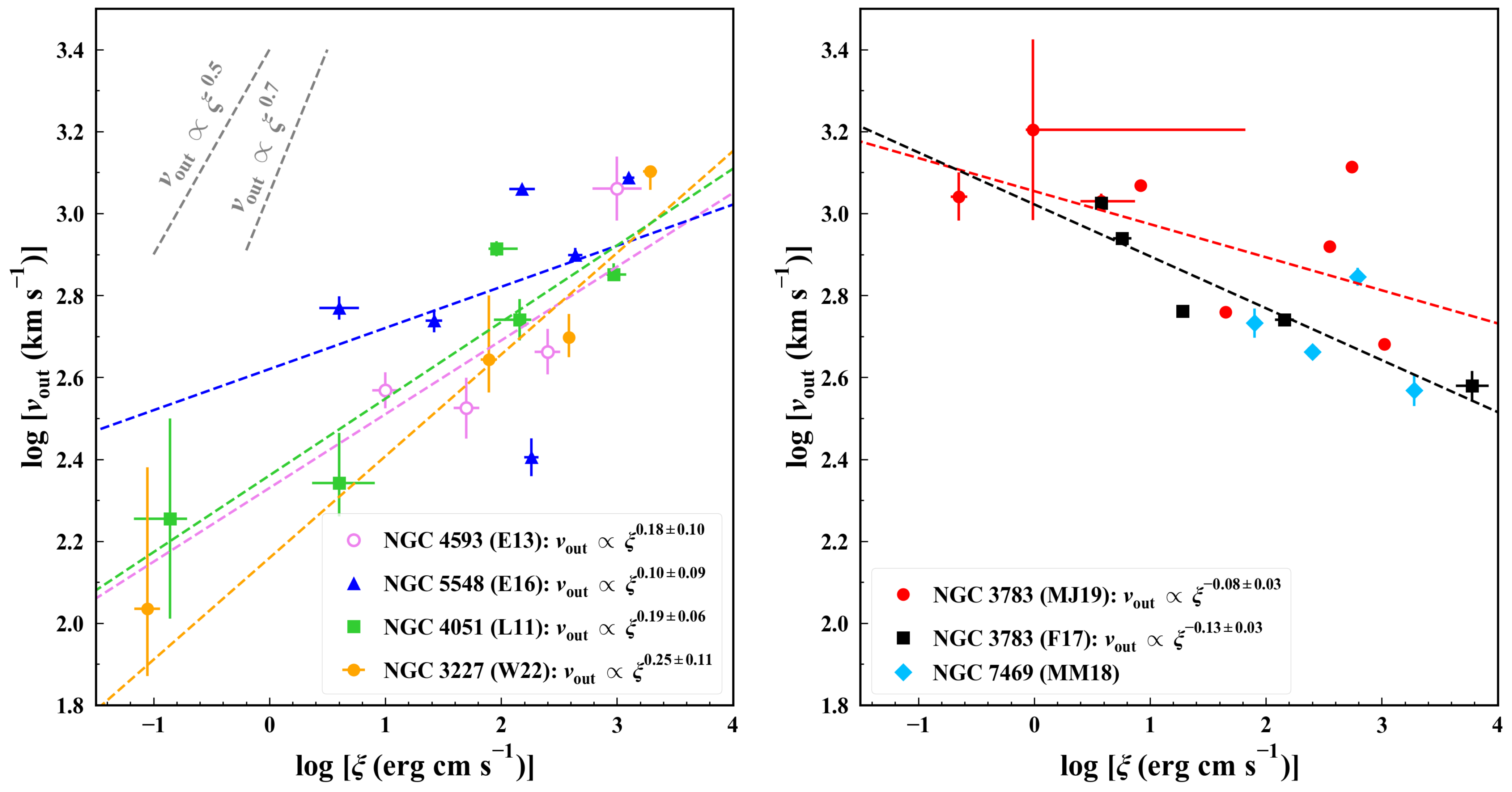}
\caption{The correlation between the outflow velocity ($v_{\rm{out}}$) and 
ionization parameter ($\xi$) for the following six Seyfert 1 galaxies:
NGC~3227 (orange solid circles in the left panel), 
NGC~3783 (red solid circles and black squares in the right panel), 
NGC~4051 (green squares in the left panel), NGC~4593 
(pink hollow circles in the left panel), NGC~5548 (blue triangles in the left panel),
and NGC~7469 (sky-blue diamonds in the right panel).
The observational data are from the previously published 
papers: \cite{Wang2022} (W22), \cite{Fu2017} (F17),
\cite{Mao2019} (MJ19), \cite{Lobban2011} (L11), 
\cite{Ebrero2013} (E13), \cite{Ebrero2016} (E16), and \cite{Mehdipour2018} (MM18).
The dashed lines represent the best-fit linear models.
The best-fit linear model for NGC~7469 cannot be constrained (see Table \ref{tab:fit}), 
so only the observational data are shown here. 
The gray dashed lines in the top left corner of the left panel represent 
the predicted correlations of radiation-driven and 
MHD-driven outflowing mechanisms.
\label{fig:general}}
\end{figure*}

   For the radiatively driven outflowing mechanism, the outflow momentum rate 
   $\dot{P}_{\rm{out}}\ (\varpropto n_{\rm{H}} r^2 v_{\rm{out}}^2)$ approximates to 
   the momentum flux of the radiation field $\dot{P}_{\rm{rad}}\ (\equiv\ L_{\rm{bol}}/c)$ \citep{Gofford2015},
   which can produce a simple scaling relation of $v_{\rm{out}}\varpropto \xi^{0.5}$ \citep{Tombesi2013}.
   For the magneto-hydrodynamically (MHD) driven outflowing mechanism,
   \cite{Fukumura2010} suggested a few scaling relations between $v_{\rm{out}}$, $r$, and $\xi$:
   $v_{\rm{out}} \varpropto r^{-\frac{1}{2}} \varpropto \xi^{\frac{1}{2(2q-1)}}$.
   \cite{Behar2009} indicated that the parameter $q$ is between 
   $\frac{6}{7}$ and 1 for WA outflowing winds in Seyfert galaxies.
   Therefore, the scaling relation between $v_{\rm{out}}$ and 
   $\xi$ in the MHD scenario is estimated to be between 
   $v_{\rm{out}} \varpropto \xi^{0.5}$ and $v_{\rm{out}} \varpropto \xi^{0.7}$ (see Figure \ref{fig:general}).
   However, the observational results show that the index of $v_{\rm{out}}$--$\xi$ relation
   is usually smaller than 0.5 
   \citep[e.g.,][]{Tombesi2013,Laha2014} or see Figure \ref{fig:general} in this work,
   which cannot be explained by the above models.

  In this work, we consider that WAs are in a pressure equilibrium state, which 
  means that the radiation pressure on the WAs is comparable to the drag
  pressure from the ACM. With that, we will use the 
  fitting results for the $v_{\rm{out}}-\xi$ relation of 
  WAs to infer the shape of the density profile of the ACM in AGNs. 
The structure of this work is shown as follows. 
The method that is applied to infer the density profile of 
the ACM in AGNs is described in Section \ref{sec:method}.
In Section \ref{sec:data}, we introduce the historical data that are used in this work. 
In Section \ref{sec:results}, we show the fitting results of the observational data,
which are further used to infer the density profile of the ACM in AGNs.
In Section \ref{sec:discussions}, we discuss the scope of application of our method.
Finally, we summarize our conclusions in Section \ref{sec:sum}.

\begin{table}[!t]
\centering
\caption{Basic properties of each object for the six Seyfert galaxies 
and previously published X-ray data used in this work. \label{tab:source}}
\begin{tabular}{llcc}
\hline\hline\xrowht[()]{30pt}
Source & \multirow{1}{1.0cm}{Seyfert type} & Redshift & WA references  \\
\hline
NGC 3227 & Sy1.5  & 0.004 & \cite{Wang2022}$^1$ \\
NGC 3783 & Sy1  & 0.010 &  \cite{Fu2017}$^1$ \\ 
                  &         &            & \cite{Mao2019}$^{1, 2}$ \\
NGC 4051 & Sy1.5  & 0.002 & \cite{Lobban2011}$^2$ \\
NGC 4593 & Sy1  & 0.008 & \cite{Ebrero2013}$^{1, 2}$ \\
NGC 5548 & Sy1.5  & 0.017 & \cite{Ebrero2016}$^{1, 2}$ \\
NGC 7469 & Sy1.2 & 0.016 & \cite{Mehdipour2018}$^2$ \\
\hline
\end{tabular}
\tablecomments{Seyfert type and redshift of each object are 
obtained from the NASA/IPAC Extragalactic Database (\href{https://ned.ipac.caltech.edu/}{NED}).
X-ray data: $^1${\it XMM-Newton}; $^2${\it Chandra}. 
}
\end{table}

\section{Method} \label{sec:method}

   The outflows in AGN might be driven by multiple mechanisms, 
   for simplicity, we only consider the radiatively driven outflowing mechanism in this work.
   The radiation pressure from the AGN radiation on the WA gas \citep{Mo2010} is
   \begin{equation}
   P_{\rm{rad}} =  \frac{L_{\rm{ion}}}{4 \pi r^2 c},
   \label{equ:rad}
   \end{equation}
   where $L_{\rm{ion}}$ is the ionizing luminosity over 1--1000 Ryd, 
   $r$ is the radial distance of the absorbing gas to 
   the central engine, and $c$ is the speed of light.
   The drag pressure \citep{Batchelor2000} produced by the ACM on the WAs is 
   \begin{equation}
   P_{\rm{D}} = \frac{1}{2} C_{\rm{D}} n_{\rm{ACM}} m_{\rm{p}} v_{\rm{out}}^2,
   \label{equ:drag}
   \end{equation}
   where $C_{\rm{D}}$ is the drag coefficient which is 
   probably equal to 1 for compressible gas or clouds, 
   $n_{\rm{ACM}}$ is the number density of the ACM,
   and $m_{\rm{p}}$ is the proton mass.
   The outflowing velocities of WAs are 
   nearly constant during several years \citep[e.g.,][]{Silva2018}. 
   In this work we assume that WAs are in a pressure equilibrium state 
   where the radiation pressure on the WAs is comparable to the drag pressure
   from the ACM:
   \begin{equation}
   P_{\rm{rad}} \simeq P_{\rm{D}}.
   \label{equ:radd}
   \end{equation}
    According to \cite{Tarter1969}, the ionization parameter of WAs can be defined by
   \begin{equation}
   \xi = \frac{L_{\rm{ion}}}{n_{\rm{e}} r^2},
   \label{equ:xi}
   \end{equation}
   where $n_{\rm{e}}$ is the electron number density 
   of the WA gas. 
   We assume that the electron number densities of the WAs gas
   and the ACM follow the power-law distributions:
   \begin{equation}
   \begin{aligned}
   n_{\rm{e}} &= n_{\rm{e,0}} \left(\frac{r}{r_0}\right)^{-m}, \\
   n_{\rm{ACM}} &= n_{\rm{ACM,0}} \left(\frac{r}{r_0}\right)^{-k},
   \end{aligned}
   \label{equ:density}
   \end{equation}
   where $r_0$ is the launching radius of the WA cloud,
   $n_{\rm{e,0}}$ is the number density of WA cloud at $r_0$ 
   and $n_{\rm{ACM,0}}$ is the number density of the ACM at $r_0$.
   Therefore, combining Equations \ref{equ:xi}--\ref{equ:density}, 
   we can obtain a correlation between $\xi$ and $v_{\rm{out}}$:
   \begin{equation}
   v_{\rm{out}} = \left[\frac{L_{\rm{ion}}^{\frac{m-k}{m-2}}}{2 \pi m_{\rm{p}} c} \cdot \frac{n_{\rm{e,0}}^{\frac{k-2}{m-2}}}{n_{\rm{ACM,0}}} \cdot r_0^{\frac{2(k-m)}{m-2}}\right]^{1/2} \xi^{\frac{k-2}{2(m-2)}}
   \label{equ:xivoutmodel}
   \end{equation}

\begin{deluxetable*}{l|l|c|c|c|c|c}
\tabletypesize{\scriptsize}
\tablewidth{0pt} 
\linespread{1.2}
\tablecaption{Best-fit parameters of $\log [v_{\rm{out}}\ ({\rm{km}}\ {\rm{s}}^{-1})] = a \times \log [\xi\ ({\rm{erg}}\ {\rm{cm}}\ {\rm{s}}^{-1})] + b$ using \textsc{LINMIX}, ODR, and BCES methods, and index $k$ of the density profile of the ACM. \label{tab:fit}}
\tablehead{
\multirow{3}{1cm}{Sources} & \multirow{3}{1cm}{Data} & \multicolumn{4}{c|}{$\log [v_{\rm{out}}\ ({\rm{km}}\ {\rm{s}}^{-1})] = a \times \log [\xi\ ({\rm{erg}}\ {\rm{cm}}\ {\rm{s}}^{-1})] + b$} & \multicolumn{1}{c}{$k=2a(m-2)+2$} \\
\cline{3-7} 
 &  & \multirow{2}{1.3cm}{Parameter} & \multicolumn{3}{c|}{Fitting method} & \multirow{2}{*}{$k\ (m=1.42)^{\textbf{\dag}}$} \\
 \cline{4-6}
 &  &                                                 & \multicolumn{1}{c|}{\textsc{LINMIX}} & \multicolumn{1}{c|}{ODR} & \multicolumn{1}{c|}{BCES} &  
}
\startdata
\multicolumn{7}{c}{Individual source}\\
\cline{1-7}
\multirow{2}{1.3cm}{NGC 3227} & \multirow{2}{0.8cm}{W22}    & $a$ & $\lesssim 0.24$  & $0.25 \pm 0.11^{\bigstar}$  & $0.35 \pm 0.10$  & $1.71\pm 0.13$ \\
\cline{3-6}
                  &                  & $b$ & $\lesssim 2.15$  & $2.16 \pm 0.29$  & $1.93 \pm 0.32$ & \\
\cline{1-7}
\multirow{2}{1.3cm}{NGC 4051} & \multirow{2}{1.2cm}{L11} & $a$ & $\ \ 0.19 \pm 0.06^{\bigstar}$  & $0.20 \pm 0.10$  & $0.03 \pm 0.09$ & $1.78\pm 0.07$ \\
\cline{3-6}
                  &                  & $b$ & $2.36 \pm 0.13$  & $2.35 \pm 0.20$  & $2.79 \pm 0.23$ & \\
\cline{1-7}
\multirow{2}{1.3cm}{NGC 4593} & \multirow{2}{0.8cm}{E13}  & $a$ & $\lesssim 0.18$  & $0.27 \pm 0.24$  & $\ \ \ \ 0.18 \pm 0.10^{\bigstar}$ & $1.79\pm 0.11$ \\
\cline{3-6}
                  &                  & $b$ & $\lesssim 2.06$  & $2.09 \pm 0.58$  & $\ \ 2.33 \pm 0.19$ & \\
\cline{1-7}
\multirow{2}{1.3cm}{NGC 5548} & \multirow{2}{0.8cm}{E16}  & $a$ & $\ \ 0.10 \pm 0.09^{\bigstar}$  & $0.13 \pm 0.18$  & $\ \ 0.24 \pm 0.12$ & $1.88\pm 0.10$ \\
\cline{3-6}
                  &                  & $b$ & $2.62 \pm 0.19$  & $2.54 \pm 0.42$  & $\ \ 2.34 \pm 0.34$ & \\
\cline{1-7}
\multirow{4}{1.3cm}{NGC 3783} & \multirow{2}{0.8cm}{F17}         & $a$ & $-0.13 \pm 0.03^{\bigstar}$ & $-0.16 \pm 0.12$ & $-0.31 \pm 0.10$ & $2.15\pm 0.03$ \\
\cline{3-6}
                  &                  & $b$ & $3.02 \pm 0.06$  & $3.04 \pm 0.11$   & $3.17 \pm 0.12$ & \\
\cline{2-7}
 & \multirow{2}{0.8cm}{MJ19}      & $a$ & $-0.08 \pm 0.03^{\bigstar}$ & $-0.10 \pm 0.07$ & $-0.08 \pm 0.09$ & $2.09\pm 0.03$ \\
\cline{3-6}
                  &                  & $b$ & $3.05 \pm 0.06$  & $3.09 \pm 0.08$  & $3.04 \pm 0.21$ & \\
\cline{1-7}
\multirow{2}{1.3cm}{NGC 7469}  & \multirow{2}{*}{MM18}  & $a$ & $\lesssim 0.64$ & $-0.06\pm 0.20$ & $0.06\pm 0.19$ & \nodata \\
\cline{3-6}
 & & $b$ & $\lesssim -0.56$ & $2.88\pm 0.55$ & $2.54\pm 0.47$ & \\
\cline{1-7}
\multicolumn{7}{c}{Total}\\
\cline{1-7}
NGC~3227 \& 4051 & W22 \& L11 \& & $a$ & $\ \ 0.19 \pm 0.02^{\bigstar}$  & $0.20 \pm 0.03$  & $0.26 \pm 0.06$ & $1.78\pm 0.02$ \\
\cline{3-6}
\& 4593 \& 5548 & E13 \& E16 & $b$ & $2.37 \pm 0.05$  & $2.36 \pm 0.07$  & $2.25 \pm 0.16$ &  \\ 
\enddata
\tablecomments{
$^{\textbf{\dag}} m=1.42$ is obtained 
for a sample of 35 Seyfert 1 galaxies from \cite{Tombesi2013} 
using the absorption measure distribution. 
The fitting results followed by ``$^{\bigstar}$'' are used to calculate the index $k$
and are plotted in Figure \ref{fig:general}.
The observational data are from the previously published 
papers: \cite{Wang2022} (W22), \cite{Fu2017} (F17),
\cite{Mao2019} (MJ19), \cite{Lobban2011} (L11), 
\cite{Ebrero2013} (E13), \cite{Ebrero2016} (E16), and \cite{Mehdipour2018} (MM18).
The data of NGC~3227, NGC~4051, NGC~4593, and NGC~5548 are also fitted 
together as a reference (see ``NGC 3227 \& 4051 \& 4593 \& 5548'' in the ``Total'').
}
\end{deluxetable*}

\begin{figure}[ht!]
\includegraphics[width=0.48\textwidth]{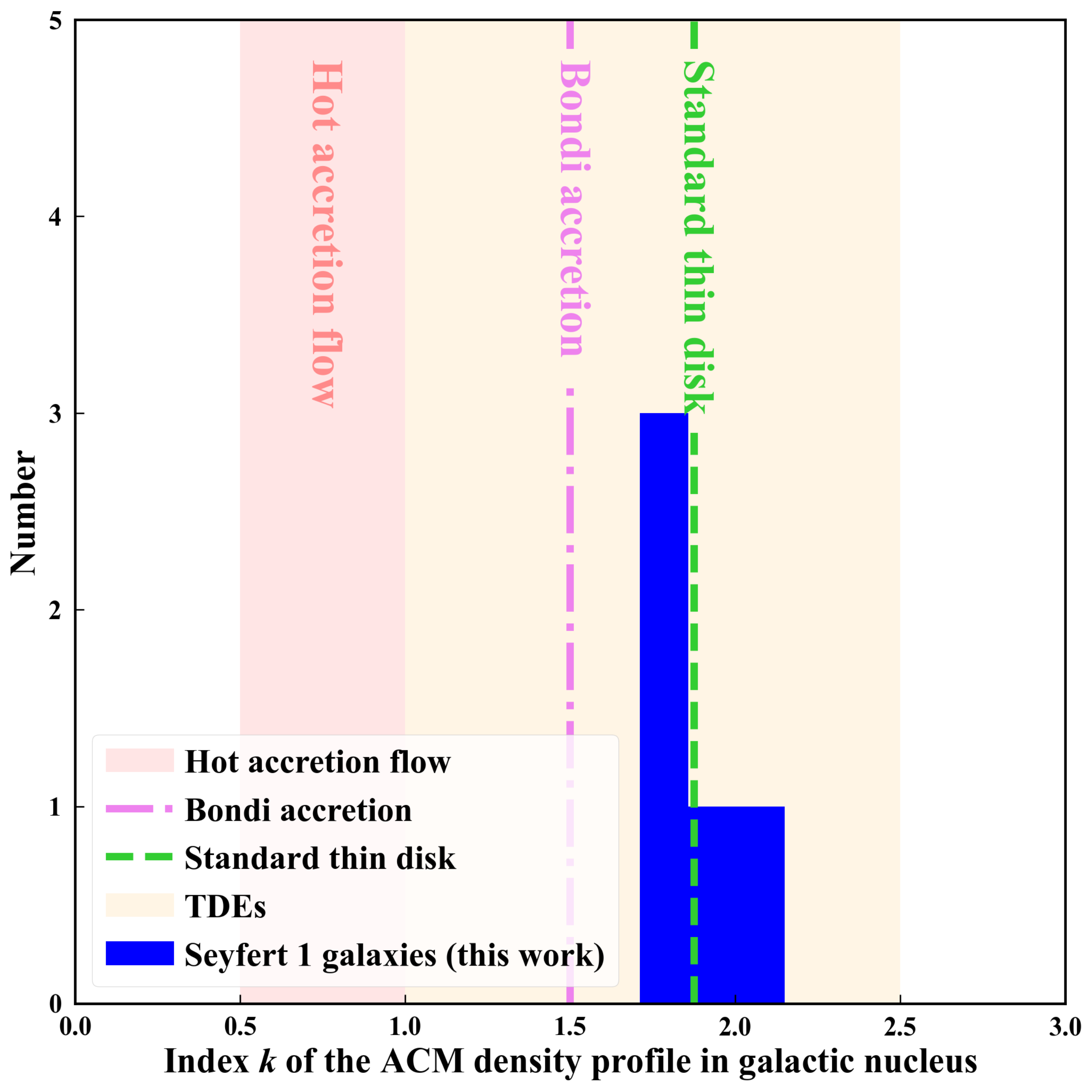}
\caption{The distribution of the ACM density profile index $k$ 
for the five Seyfert 1 galaxies (NGC~3227, NGC~4051, NGC~4593, 
NGC~5548, and NGC~3783) in this work (blue histogram).
The green dashed line represents the predicted index by the standard thin disk model \citep{Frank2002}.
The purple dash-dotted line represents the predicted index by the Bondi accretion model
\citep{Frank2002}.
The red region represents the predicted range of the index by the hot accretion flow simulations \citep{Yuan2012}.
The yellow region represents the observational range of the index in TDEs \citep{Alexander2020}.
\label{fig:kdis}}
\end{figure}

\section{Data and fitting} \label{sec:data}

In order to describe the correlation between $v_{\rm{out}}$ and $\xi$ 
of WAs for the observational data in individual AGN (see Equation \ref{equ:xivoutmodel}),
high-resolution X-ray spectra and at least four WA components are required.
Finally, we collect the parameters of WAs from the previously published papers
for the following six Seyfert 1 galaxies (see Table \ref{tab:source}): 
\begin{itemize}[leftmargin=*]
\item[\textbf{--}] NGC~3227: \cite{Wang2022} found four 
WA components using the {\it XMM-Newton} spectra data.
\item[\textbf{--}] NGC~3783: \cite{Fu2017} found five WA components 
through fitting the {\it XMM-Newton} spectra,
while \cite{Mao2019} found nine WA components 
using both the {\it XMM-Newton} and {\it Chandra} data.
\item[\textbf{--}] NGC~4051: \cite{Lobban2011} found five WA components 
using the {\it Chandra} spectral data.
\item[\textbf{--}] NGC~4593: \cite{Ebrero2013} found four WA components
through fitting the spectra of {\it XMM-Newton} and {\it Chandra}.
\item[\textbf{--}] NGC~5548: \cite{Ebrero2016} found six WA components 
through fitting the spectra of {\it XMM-Newton} and {\it Chandra} \citep[H02 data in][]{Ebrero2016}.
\item[\textbf{--}] NGC~7469: \cite{Mehdipour2018} found four WA components 
through fitting the spectra of {\it Chandra}.
\end{itemize}
Then we fit the correlation between $\xi$ and $v_{\rm{out}}$ in each source using the following linear model:
\begin{equation}
\log [v_{\rm{out}}\ ({\rm{km}}\ {\rm{s}}^{-1})] = a \times \log [\xi\ ({\rm{erg}}\ {\rm{cm}}\ {\rm{s}}^{-1})] + b,
\label{equ:xivoutdata}
\end{equation}
where $a$ corresponds to the theoretical index $(k-2)/[2(m-2)]$ in Equation \ref{equ:xivoutmodel},
i.e., $a=(k-2)/[2(m-2)]$. 
Therefore, the index of the density profile of the ACM can be calculated by 
\begin{equation}
k=2a(m-2)+2.
\label{equ:ISMk}
\end{equation}
We mainly use \textsc{LINMIX}\footnote{https://linmix.readthedocs.io/en/latest/src/linmix.html} method 
\citep[][]{Kelly2007} to fit the observational data.
The \textsc{LINMIX} method performs the linear regression 
based on a Bayesian approach, 
which runs a Markov-chain-Monte-Carlo algorithm to calculate the posterior distribution
and can account for measurement errors on both variables in the fit.
However, for NGC~3227 and NGC~4593, 
this method can only give an upper limit for the parameters (see Table \ref{tab:fit}). 
Therefore, we also use the following two methods as supplements: 
Orthogonal Distance Regression\footnote{https://docs.scipy.org/doc/scipy/reference/odr.html} 
\citep[ODR;][]{Boggs1989}, 
and bivariate correlated errors and intrinsic 
scatter\footnote{https://github.com/rsnemmen/BCES} \citep[BCES;][]{Akritas1996,Nemmen2012}.
Both of these two methods can also deal with measurements errors on both variables.
The BCES method is a weighted least squares estimator,
and the ODR method uses the least
squares method to minimize the
weighed orthogonal distance from the data to the fitted curve.
The \textsc{LINMIX} method can provide a consistent fitting result to 
at least one of the other two methods (see Table \ref{tab:fit}).

\begin{figure*}[ht!]
\centering
\includegraphics[width=\linewidth, clip]{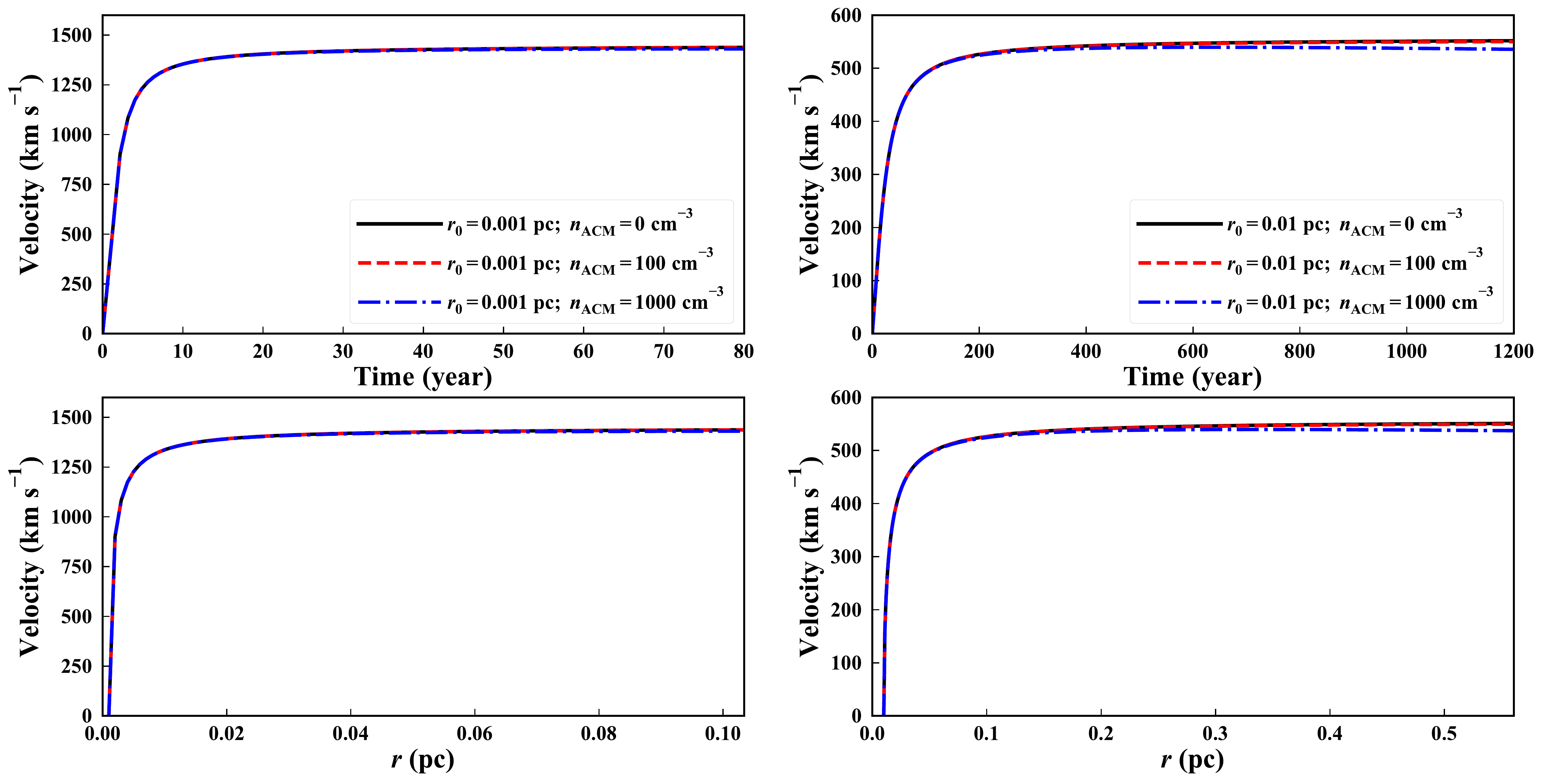}
\caption{Estimating the acceleration timescale of WA outflows.
The left two panels show the case of the launching radius ($r_0$)
of 0.001 pc and the right two panels show the case of the launching radius of 0.01 pc. 
The black solid, red dashed and blue dash-dotted lines represent the calculation of 
the ACM number density ($n_{\rm{ACM}}$) of 0 cm$^{-3}$, 100 cm$^{-3}$, and 1000 cm$^{-3}$, respectively.
It is obvious that the acceleration timescale is much shorter than the lifetime of WA.
\label{fig:model}}
\end{figure*}


\section{Results}
\label{sec:results}

As Figure \ref{fig:general} shows, there is a positive correlation between
$v_{\rm{out}}$ and $\xi$ for NGC~3227, NGC~4051, NGC~4593, and 
NGC~5548 (The coefficient $a$ of Equation \ref{equ:xivoutdata} 
ranges from 0.10 to 0.25; also see Table \ref{tab:fit}), 
while NGC~3783 shows a negative correlation.
For NGC~3783, $a$ is $-0.13\pm 0.03$ for the data from \cite{Fu2017},
and is $-0.08\pm 0.03$ for the data from \cite{Mao2019}.
However, the error bar of coefficient $a$ is large, so we fit the 
data of NGC~3227, NGC~4051, NGC~4593, and NGC~5548 together as a reference, 
resulting in $a=0.19\pm 0.02$.
The fitting result in individual source is consistent 
with the total fitting result in the sample.
As mentioned in Section \ref{sec:intro}, the power-law indexes $a$ for these 
Seyfert 1 galaxies are smaller than the predicted values by the theories: 
0.5 for the radiatively driven outflowing mechanism, 
and larger than 0.5 for the MHD driven outflowing mechanism.
The $v_{\rm{out}}-\xi$ relation of NGC~7469 cannot be constrained,
so its ACM density profile will not be discussed further (see Table \ref{tab:fit}),
and its observational data are shown in Figure \ref{fig:general} as a reference.

The number density distribution of WAs can be estimated by the 
absorption measure distribution \citep{Holczer2007,Behar2009}.
\cite{Tombesi2013} estimated that $m=1.42$ for WAs in a sample of 35 Seyfert 1 galaxies. 
Combining Equation \ref{equ:ISMk} and $m=1.42$ \citep{Tombesi2013}, 
the density profiles of the ACM in these 
Seyfert 1 galaxies are estimated to be between 
$n \propto r^{-1.7}$ and $n \propto r^{-2.15}$ (see Table \ref{tab:fit})
from 0.01 pc to pc scales, or even larger scales (the distance range of WAs).
Both \cite{Tombesi2013} and \cite{Laha2014} investigated the correlation between 
$v_{\rm{out}}$ and $\xi$ for WAs in a large AGN sample,
which obtained $a=0.31$ and $a=0.12$, respectively.
Therefore, index of $k$ is 1.64 and 1.82 for 
\cite{Tombesi2013} and \cite{Laha2014}, respectively.
Our results are similar to those in AGN samples.
The density profile indexes $k$ of the ACM in the five Seyfert 1 galaxies of our sample
(NGC~3227, NGC~3783, NGC~4051, NGC~4593, and NGC~5548)
are within the range of $k$ for the ACM in TDEs 
\citep[between $-1$ and $-2.5$;][]{Alexander2020} (see Figure \ref{fig:kdis}).

The density profile of the ACM within the Bondi radius might be
connected to the accretion models. The Bondi radius can be expressed by
$r_{\rm{B}} = 2GM_{\rm{BH}}/c_{s, \infty}^2$,
where $M_{\rm{BH}}$ is the SMBH mass and $c_{\rm{s, \infty}}$ 
is the sound speed at infinity \citep{Bondi1952}.
For simplicity, we assume that the sound speeds at infinity of our sample 
are similar to that of M87 
\citep[$r_{\rm{B}}=$ 0.11--0.22 kpc with $M_{\rm{BH}}=3.5\times 10^9\ M_\odot$;][]{Russell2015} 
and Sgr A* \citep[$r_{\rm{B}}=$ 0.4 pc with $M_{\rm{BH}}=4\times 10^6\ M_\odot$;][]{Li2015}.
Thus, according to the average $M_{\rm{BH}}$ of our sample 
\citep[$\sim$ $10^7$ $M_\odot$;][]{Bentz2015},
the Bondi radii of our sample might be between 0.5 pc and 1 pc.
Warm absorbers can exist from the scale within the Bondi radius 
\citep[e.g.,][]{Ebrero2016,Wang2022} 
to the kpc scale \citep{Laha2021}.
Although the large scale might not be associated with the accretion flow,
given that most of the WAs in our sample might be located within 
or around the Bondi radius \citep[e.g.,][]{Ebrero2016,Wang2022},
we can briefly compare the density profiles between the ACM and the accretion flow here.
The indexes $k$ of the five Seyfert 1 galaxies are  
larger than the predicted value by 
the spherically symmetrical Bondi accretion model 
\citep[$-1.5$;][]{Frank2002} and the simulated results of the 
hot accretion flow \citep[between $-0.5$ and $-1.0$;][]{Yuan2012},
but relatively consistent with the prediction by the 
standard thin disk model \citep[$-15/8$;][]{Frank2002}
(see Figure \ref{fig:kdis}).

\section{Discussions}
\label{sec:discussions}
\subsection{Acceleration timescale required before equilibrium}
\label{sec:timescale}
To verify whether the assumption about the pressure equilibrium is feasible,
we firstly estimate the acceleration timescale before reaching equilibrium of WA outflows. 
Under the action of the radiation pressure and drag pressure, 
the motion equation of the WA clouds is
\begin{equation}
\frac{vdv}{dr} = \frac{f_{L}L_{\rm{ion}}}{4\pi c N_{\rm{H}} m_{\rm{p}} r^2}- \frac{C_{\rm{D}} n_{\rm{ACM}}}{2 N_{\rm{H}}} v^2,
\end{equation}
where $f_{L}$ is the fraction of the ionizing luminosity being absorbed or scattered by the WA cloud,
which is about 2\% according to \cite{Grafton2020} and \cite{Wang2022} and
$m_{\rm{p}}$ is the mass of proton.
The average ionizing luminosity of the sources in our sample is $5\times10^{43}\ {\rm{erg}}\ {\rm{s}}^{-1}$.
We simply set a constant column density to be $N_{\rm{H}}$=$10^{22.5}\ {\rm{cm}}^{-2}$,
which is the maximum $N_{\rm{H}}$ for WAs obtained in AGN samples \citep{Tombesi2013,Laha2014}.
As shown in Figure \ref{fig:model}, we calculate the acceleration timescale for the launching radii
$r_0$ of 0.001 pc and 0.01 pc, with $n_{\rm{ACM}}$ being 0 cm$^{-3}$, 100 cm$^{-3}$, and 1000 cm$^{-3}$.
For the WA component that is close to the SMBH, 
the typical acceleration distance might be about 0.01 pc 
and the typical acceleration timescale might be about 10 years 
(see the left two panels of Figure \ref{fig:model}), 
while the existence distance of this WA component 
might be larger than 0.01 pc \citep{Laha2021}, 
which means that its existence timescale might be longer than its acceleration timescale.
For the WA component that is relatively farther,
the typical acceleration distance might be about 0.05 pc
and the typical acceleration timescale might be about 100 years
(see the right two panels of Figure \ref{fig:model}),
while the existence distance of this WA component is larger than 0.05 pc
\citep[e.g.,][]{Ebrero2016,Wang2022},
which indicates that its acceleration timescale might be shorter than the existence timescale.
These imply that the lifetimes of WAs are much larger than the 
acceleration timescales. These results indicate that WAs can stay 
in an equilibrium state during the most periods of their life.

\subsection{Imbalance caused by AGN variabilities}
The AGN variabilities can break the equilibrium state of WAs.
Assuming the central luminosity changes from $L$ to $L'$ with $L'=(1+f)L$, the radiation pressure acting on each 
component of WAs along the line of sight will become $P_{\rm{rad}}'=(1+f)P_{\rm{rad}}$ one by one.
According to Equations. \ref{equ:rad} and \ref{equ:drag},
then we can easily find that $P_{\rm{rad}}'=(1+f)P_{\rm{D}}$ for each component of WAs.
This means that the variability only has an impact on the estimation for the coefficient
of Eq. \ref{equ:xivoutmodel} rather than the index.
That is to say, even if the AGN variabilities are considered, the estimation for the index
of the ACM density profile will not be affected.

\section{Summary}
\label{sec:sum}

In this work, we propose a novel method to measure the ACM density profile
by the equilibrium between the radiation pressure on the WA outflows
and the drag pressure from the ACM for the following six Seyfert 1 galaxies:
NGC~3227, NGC~3783, NGC~4051, NGC~4593, NGC~5548, and NGC~7469.

We study the correlation between outflow 
velocity and ionization parameter of the WAs in 
five Seyfert 1 galaxies of our sample 
(NGC~3227, NGC~3783, NGC~4051, NGC~4593, and NGC~5548).
According to the fitting results of the $v_{\rm{out}}-\xi$ relation, 
we infer that the density profile of 
the ACM is between $n \propto r^{-1.7}$ and 
$n \propto r^{-2.15}$ from 0.01 pc to pc scales in these five AGNs. 
The indexes of the ACM density profiles in these five 
Seyfert galaxies are within the range of the indexes in TDEs.
Our results indicate that
the ACM density profile in Seyfert 1 galaxies is steeper than the prediction by the 
spherically symmetric Bondi accretion model and the simulation results of 
the hot accretion flow, but more in line with the prediction by the standard thin disk model.

\begin{acknowledgments}
We thank the referee for constructive comments that improved this paper.
Y.J.W. acknowledges support from the start-up research fund of 
School of Astronomy and Space Science of Nanjing University. 
Y.J.W. and Y.Q.X. acknowledge support from NSFC-12025303, 11890693, the CAS Frontier Science Key Research Program 
(QYZDJ-SSW-SLH006), and the K.C. Wong Education Foundation.
Z.-C. H. is supported by NSFC-11903031, 12192220, 12192221 and USTC Research 
Funds of the Double First-Class Initiative YD 3440002001.
This research has made use of the NASA/IPAC Extragalactic Database (NED), 
which is funded by the National Aeronautics and Space Administration 
and operated by the California Institute of Technology.
\end{acknowledgments}

\bibliography{ms_ACM}{}
\bibliographystyle{aasjournal}

\end{CJK*}
\end{document}